\documentclass[10pt]{article}
\usepackage{geometry}                
\usepackage{graphicx}
\usepackage{hyper ref}
\usepackage{placeins}
\usepackage{amssymb}
\usepackage{epstopdf}
\usepackage{authblk}
\begin{document}

\title{Highlights from the Compact Muon Solenoid (CMS)~Experiment$\dagger$}
\author[1,*]{Saranya Samik Ghosh}
\author[]{on behalf of the CMS Collaboration}

\affil[1]{III. Physikalisches Institut A,
Physikzentrum, RWTH Aachen University, 52062 Aachen, Germany}

\date{October 2018}

\affil[*]{saranya.ghosh@cern.ch}

\maketitle

 \footnote{\hangafter=1 \hangindent=1.05em \hspace{-0.82em} $\dagger$ This paper is based on the talk at the 7th International Conference on New Frontiers in Physics (ICNFP 2018), Crete, Greece, 4-12 July 2018.} 

\abstract{ 
The highlights of the recent activities and physics results leading up to the summer of 2018 from the Compact Muon Solenoid (CMS) experiment at the CERN Large Hadron Collider (LHC) are presented here. The CMS experiment has a very wide-ranging physics program, and only a very limited subset of the physics analyses being performed at CMS are discussed here, consisting of several important results from the analysis of proton-proton collision data at center-of-mass energy of 13 TeV. These include important analyses of Higgs boson physics, with the highlight being the first observation of the $\mathrm{t\bar{t}H}$ production of the Higgs boson, along with analyses pertaining to precision standard model measurements, top quark physics, with the single top production cross-section measurement,  and flavor physics, with the important observation of $\chi_{b}$(3P) states. Additionally, important searches for physics beyond the standard model are also presented.
}


\newpage


\section{Introduction}\label{sec:intro}
The Compact Muon Solenoid (CMS) experiment \cite{bib:CMS} is one of the four large experiments at the CERN Large Hadron Collider (LHC) with a wide range of physics goals. The CMS experiment had a very successful physics program during the first run (Run 1) of the LHC between 2010 and 2012 with the datasets collected with proton-proton collisions at center-of-mass energy of 7 and 8 TeV, with the highlight being the discovery of a new particle with a mass of around 125 GeV \cite{bib:higgs_CMS} (also by the ATLAS experiment \cite{bib:higgs_ATLAS}) that is consistent with the hypothesized Higgs boson, the particle associated with the Brout-Englert-Higgs mechanism for electroweak symmetry breaking \cite{bib:higgs_theory1, bib:higgs_theory2, bib:higgs_theory3, bib:higgs_theory4, bib:higgs_theory5, bib:higgs_theory6}. Along with the studies related to the Higgs boson, CMS has a diverse physics program covering, but not limited to, precision standard model (SM) measurements, studies with the top quark, flavor physics, and a wide range of searches for physics beyond the standard model (BSM). 

The CMS experiment continues its diverse physics program at the second run of the LHC (Run~2) with proton-proton collisions at center-of-mass energy of 13 TeV from 2015 through to 2018. After~the discovery of a particle consistent with the Higgs boson (hereafter referred to simply as the Higgs boson), the measurements of the properties of the newly discovered particle has become one of the most important areas of study. The first observation of the $\mathrm{t\bar{t}H}$ production mechanism of the Higgs boson reported in 2018 by the CMS \cite{bib:cms_tth_comb} experiment has been the highlight in this area. The increase in the center-of-mass energy of the proton-proton collisions also provides access to wider regions in the phase space for searches for physics BSM. Precision SM measurements, measurements of flavor physics and top quark physics also gain from a larger dataset collected at 13 TeV. 

This document presents some recent important results from physics analyses reported by the CMS experiment. The results presented here constitute only a small subset of the physics analyses at CMS. Firstly, the CMS experiment is described in Section~\ref{sec:cms} with a brief description of the CMS detector in Section~\ref{subsec:cms_detector} followed by a short discussion in the performance of the CMS detector during LHC Run 2 in Section~\ref{subsec:cms_performance}. Then, the recent physics highlights from CMS are presented in Section~\ref{sec:physics_results}, with the results from Higgs boson physics presented in Section~\ref{subsec:higgs}, precision SM measurements described in Section~\ref{subsec:standard_model}, results from top physics in Section~\ref{subsec:top}, important measurements from flavor physics in Section~\ref{subsec:flavour} and the searches for new physics BSM in Section~\ref{subsec:bsm}. Finally, Section~\ref{sec:conclusions} contains the concluding remarks.

\section{CMS Experiment at the LHC}
\label{sec:cms}

The CMS experiment \cite{bib:CMS} is one of the four large experiments at the LHC. A very brief description of the CMS detector along with a short discussion about its performance can be found below.

\subsection{The CMS Detector}
\label{subsec:cms_detector}
\begin{figure}[htbp]
\begin{center}
\includegraphics[width=12.5cm,clip]{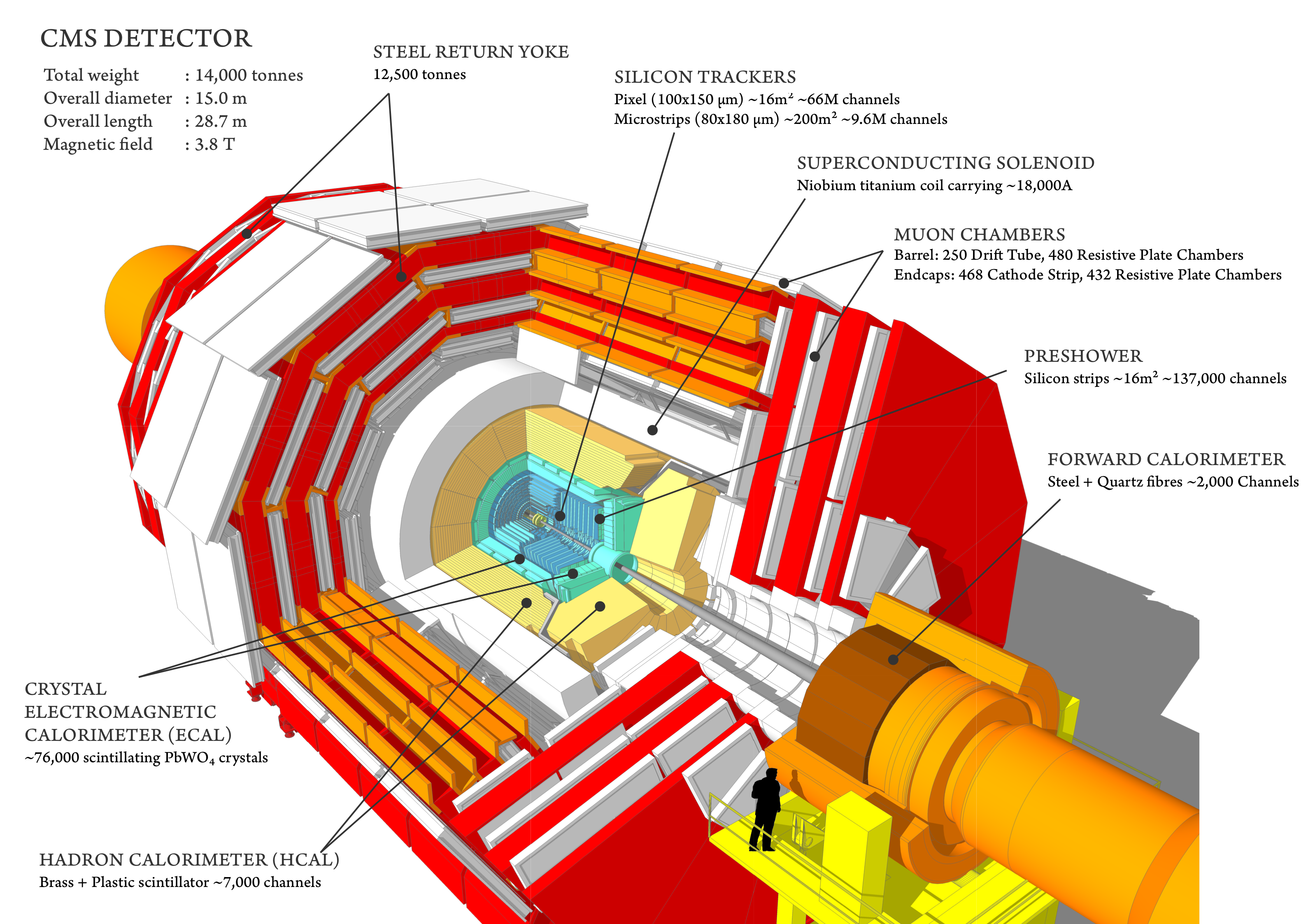}
\caption{An overview of the CMS detector.}
\label{fig:CMS}       
\end{center}
\end{figure}

The CMS detector is a general-purpose detector at the LHC. An overview of the different parts of the CMS detector can be found in Figure~\ref{fig:CMS}. The central feature of the CMS detector is a superconducting solenoid magnet that is operating under conditions to provide a magnetic field of 3.8 T. Inside the solenoid, the first sub-detector closest to the point of the collisions is the silicon pixel detector, followed by a silicon strip tracker, with both together forming the tracker system of CMS. Outside the tracker system is the electromagnetic calorimeter that is made up of lead tungstate crystals, followed by the hadronic calorimeter, which is a sampling calorimeter. Outside the solenoid magnet, embedded with the iron return yoke, is the muon system consisting of drift tubes, cathode strip chambers and resistive plate chambers. Additionally, two hadronic forward calorimeters are positioned at either end of CMS to measure particles with a trajectory close to the beam pipe.
Further details about the CMS detector can be found in Ref.~\cite{bib:CMS}.

\FloatBarrier

\subsection{CMS Performance in LHC Run 2}
\label{subsec:cms_performance}

\begin{figure}[htbp]
\begin{center}
\includegraphics[width=7cm,clip]{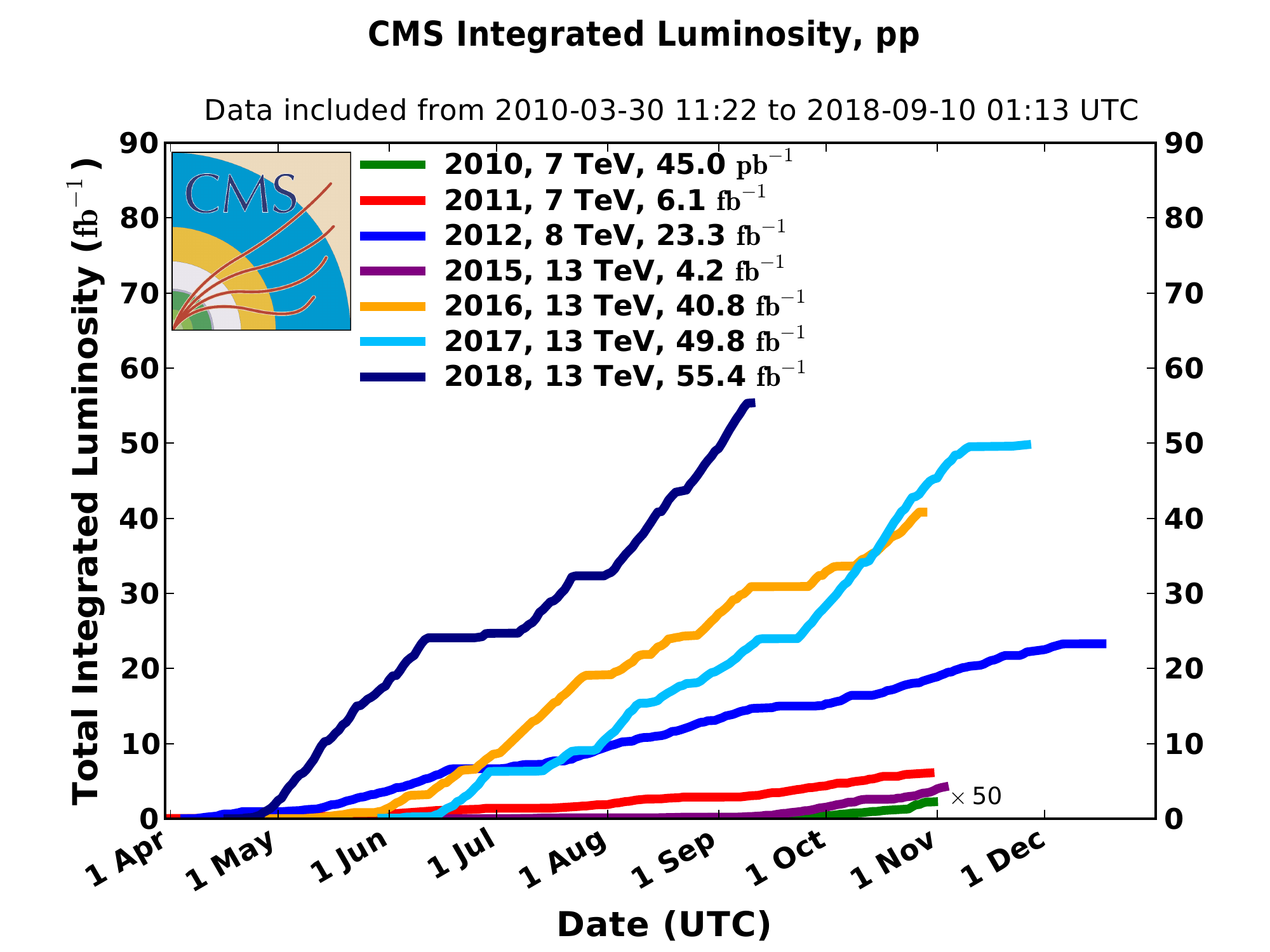}
\includegraphics[width=7cm,clip]{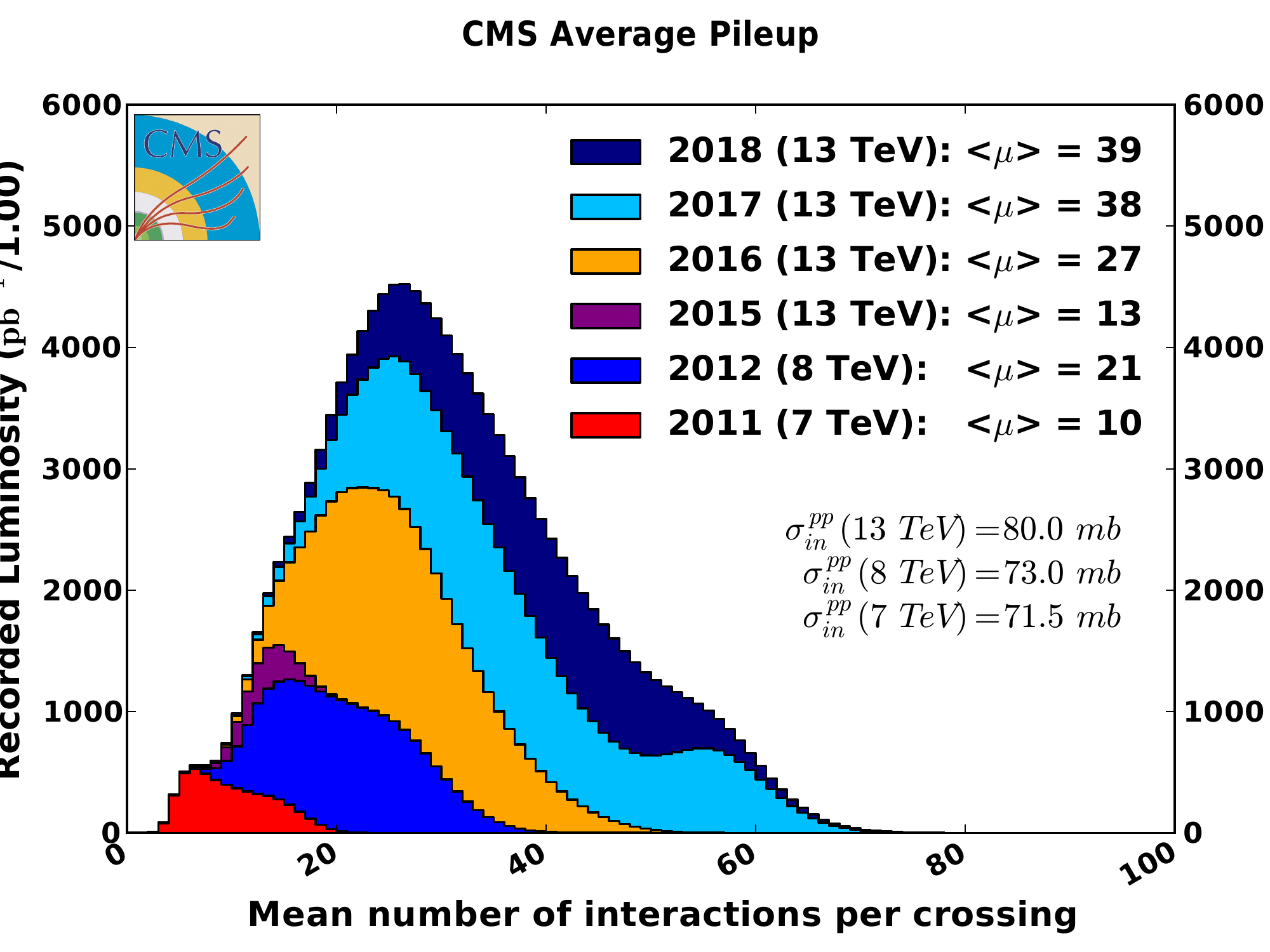}
\caption{The total integrated luminosity (\textbf{left}) and the mean number of interactions per bunch crossing (\textbf{right}) for the datasets collected during proton-proton collisions by the CMS experiment in the different years of operation \cite{bib:cms_lumi}.}
\label{fig:cms_lumi}       
\end{center}
\end{figure}

The CMS experiment has been performing very well, while recording and analyzing the data produced by the LHC with high efficiency. After the Run 1 of the LHC (the period of running between 2010 and 2012), there was a period of two years of long shutdown when certain sub-detectors at CMS were upgraded, with the pixel detector upgrade being the most prominent of these. The~Run~2 of the LHC started in 2015 and continues through to 2018, with proton-proton collisions at center-of-mass energy of 13 TeV and higher instantaneous luminosity than ever before, reaching over 21 Hz/nb. The~high instantaneous luminosity presents its own challenges, producing an average of 37 interactions per bunch crossing during the proton-proton collision run in 2017. CMS has been able to record the data and reconstruct it for use in physics analysis with high efficiency, with a total integrated luminosity of 37.76 $\mathrm{fb^{-1}}$ recorded out of 40.82 $\mathrm{fb^{-1}}$ delivered by the LHC during 2016, and 44.98 $\mathrm{fb^{-1}}$ recorded out of 49.79 $\mathrm{fb^{-1}}$ delivered during 2017. Of the dataset recorded, a very high fraction of it has been certified as being good for physics analysis where all the relevant detector components were checked to be performing satisfactorily. An even higher amount of data is expected to be collected by the end of 2018. The progression of the amount of data collected by the CMS experiment and the mean number of interactions per bunch crossing of the collisions, which is directly correlated with the instantaneous luminosity, can be seen in Figure~\ref{fig:cms_lumi}.

\section{Physics Highlights from CMS}
\label{sec:physics_results}

Selected important recent physics results from CMS are presented in the following sections.
A~complete list of publications of the CMS collaboration can be found in Ref.~\cite{bib:cms_public_web}

\subsection{Higgs Boson Physics}
\label{subsec:higgs}

After the discovery of the Higgs boson, measurements of its properties and couplings to other particles have become important goals of the CMS experiment. In pursuit of that goal, probes of the $\mathrm{t\bar{t}H}$ production mechanism and the search for di-Higgs production conducted at CMS are presented~below.

\subsubsection{Observation of the $\mathrm{t\bar{t}H}$ Process}

\begin{figure}[htbp]
\begin{center}
\includegraphics[width=7.1cm,clip]{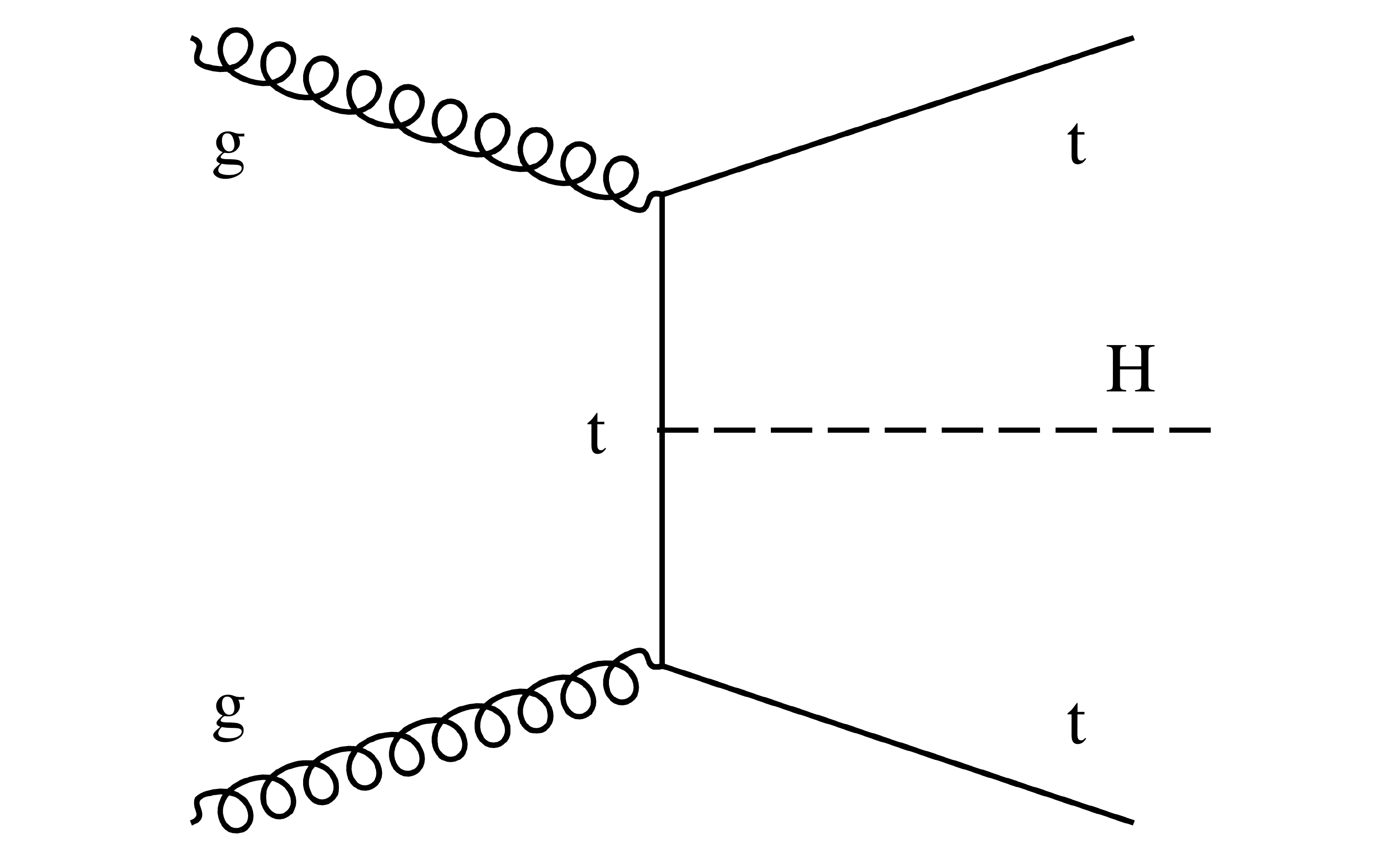}
\includegraphics[width=5.6cm,clip]{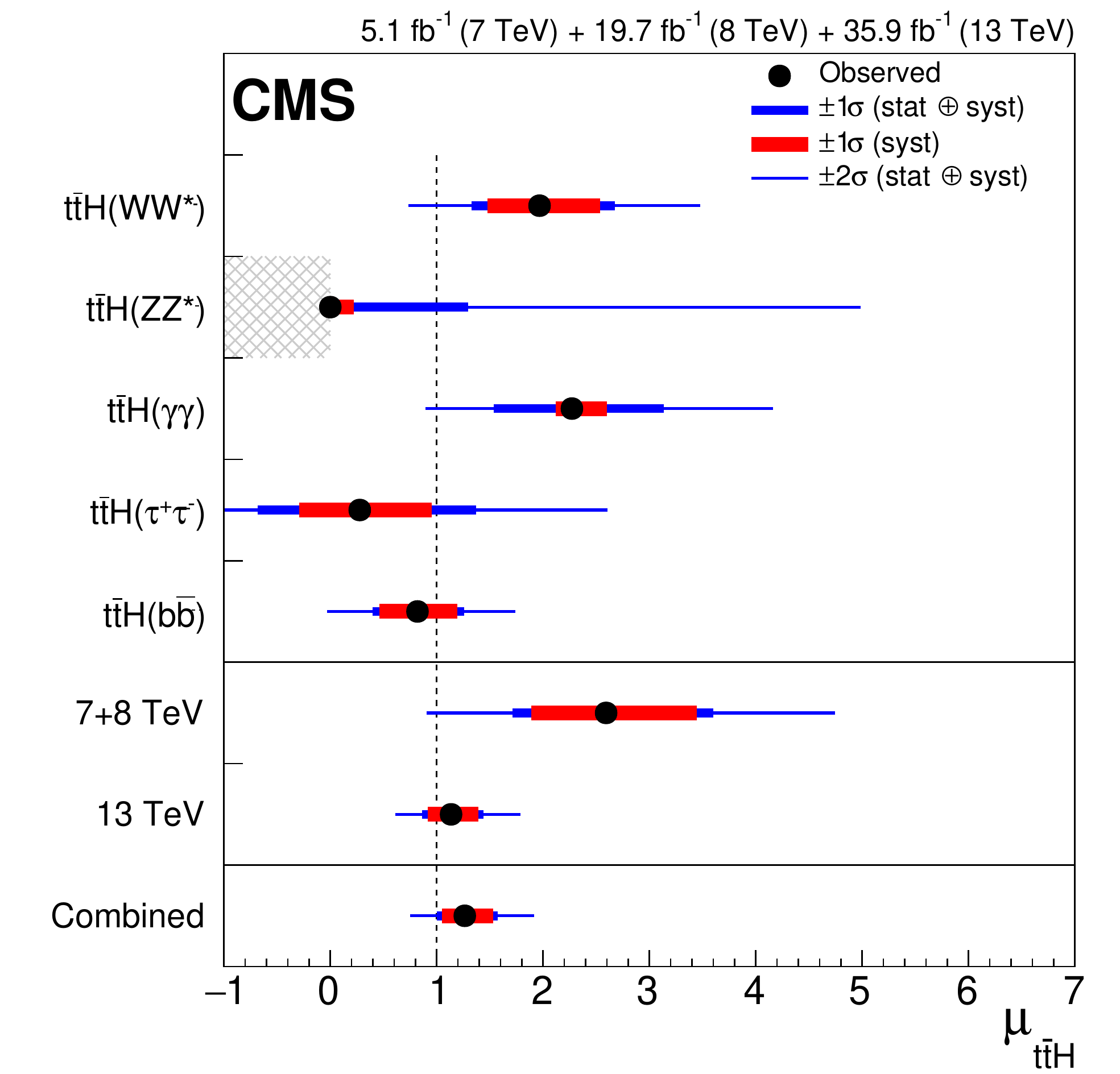}
\caption{The Feynman diagram of the $\mathrm{t\bar{t}H}$ process (\textbf{left}) and combination of the measured signal strengths from the different channels of search for the $\mathrm{t\bar{t}H}$ process (\textbf{right}) \cite{bib:cms_tth_comb}.}
\label{fig:tth_comb}       
\end{center}
\end{figure}

 The associated production of the Higgs boson with a pair of top quarks ($\mathrm{t\bar{t}H}$) gives direct access to the top quark Yukawa coupling, since it probes the coupling of the Higgs boson to the top quark directly at tree level. It is an important process that could provide access to physics BSM if the coupling is found to be inconsistent with the SM expectation. The~Feynman diagram of the $\mathrm{t\bar{t}H}$ process is shown in Figure~\ref{fig:tth_comb}(left). 

In CMS, there are several different analyses probing the $\mathrm{t\bar{t}H}$ process in the different final states corresponding to the $\mathrm{b\bar{b}}$, multi-lepton (corresponding to ZZ*, WW* and $\tau\tau$) and diphoton decay channels of the Higgs boson \cite{bib:tth_cms_bb_jet, bib:tth_cms_bb_lep, bib:tth_cms_multilepton, bib:tth_cms_4l, bib:tth_cms_yy, bib:tth_cms_Run1}. A combination of the measurements made in the different channels at 13 TeV, 8 TeV and 7 TeV has been performed, leading to the first experimental observation of the $\mathrm{t\bar{t}H}$ process with a significance of 5.2 standard deviations over the expectation from the background-only hypothesis  \cite{bib:cms_tth_comb}. The signal strength parameter, corresponding to the ratio of the observed cross-section to the value of the cross-section predicted by the SM, has been measured to be $\mathrm{1.26^{+0.31}_{-0.26}}$, which is consistent with the SM prediction. The combination of the measured signal strengths from the different channels is shown in Figure~\ref{fig:tth_comb}(right).

\subsubsection{Search for Higgs Boson Pair Production }

\begin{figure}[htbp]
\begin{center}
\includegraphics[width=7cm,clip]{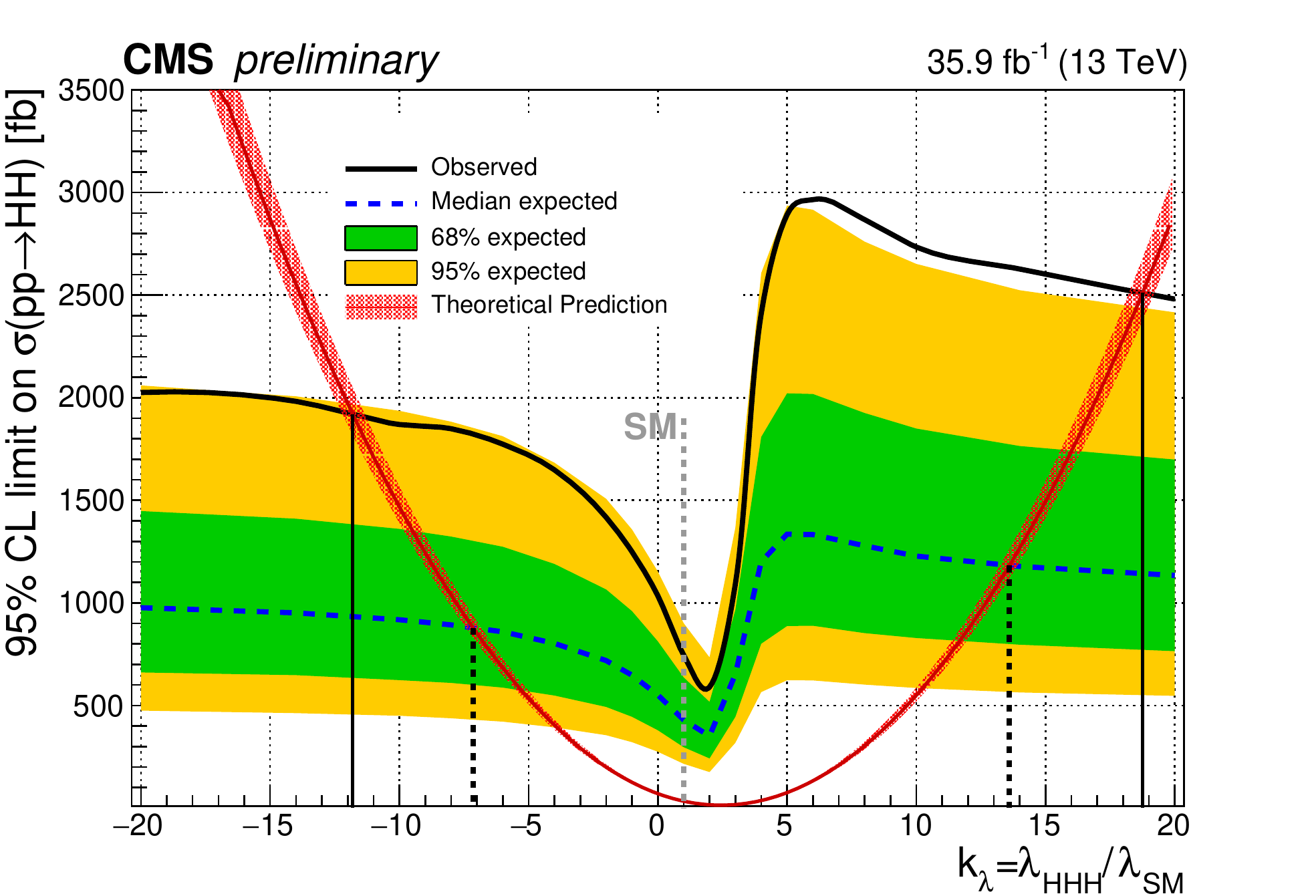}
\includegraphics[width=7cm,clip]{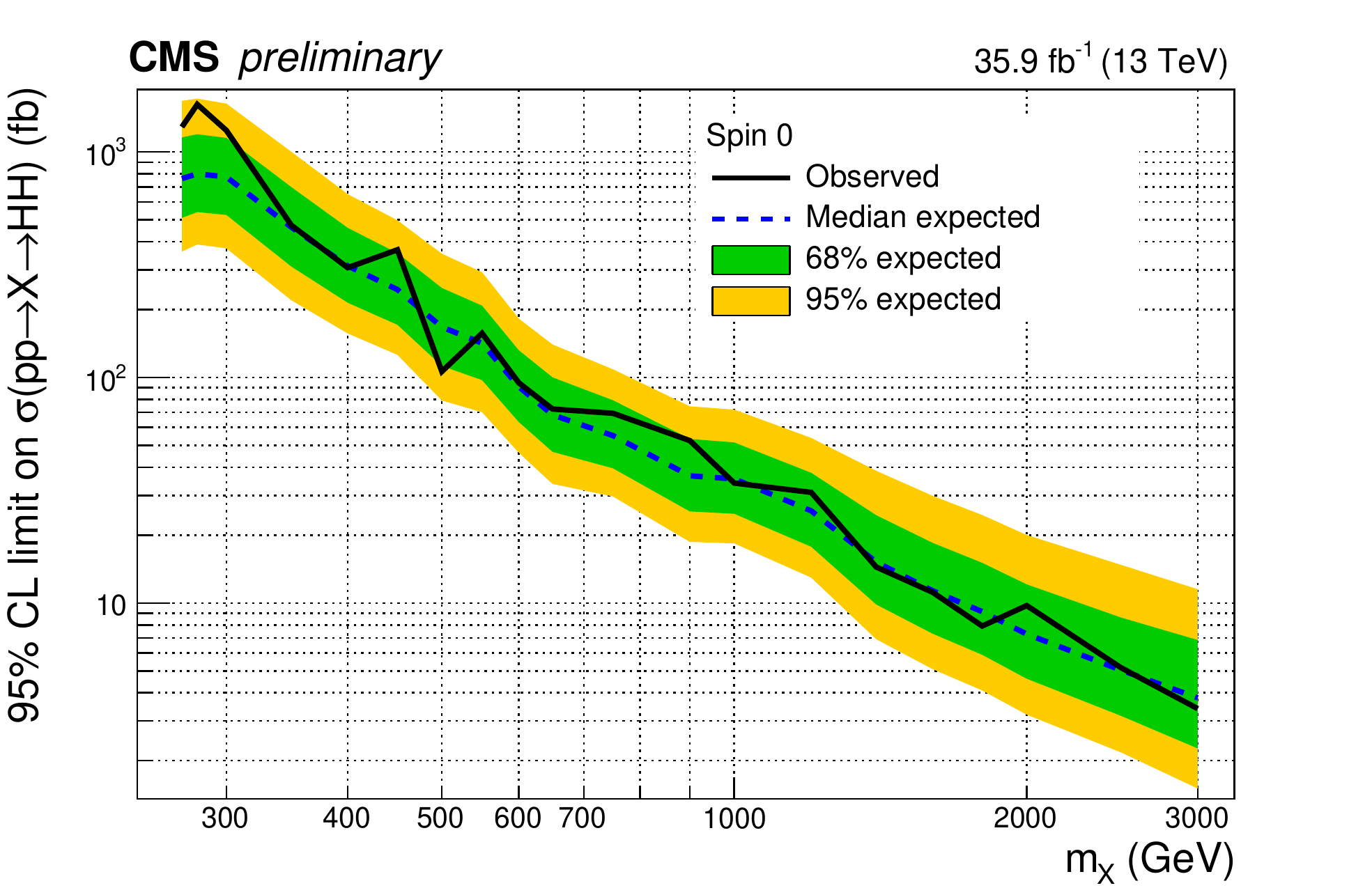}
\caption{Expected and observed 95\% CL upper limits on the non-resonant HH production cross-section as a function of the $k_{\lambda}$ parameter, that affects both the expected cross-section and the HH decay kinematics, as described in Ref.~\cite{bib:cms_HH} (\textbf{left}) and the  exclusion limits on the production of a narrow width spin zero resonance decaying into a pair of Higgs bosons (\textbf{right}) \cite{bib:cms_HH}. }
\label{fig:hh}       
\end{center}
\end{figure}

A search for the Higgs boson pair production (HH), which probes the self-coupling of the Higgs boson, has also been performed at CMS \cite{bib:cms_HH}. The search has been performed in different decay channels of the Higgs boson, corresponding to the channels with one Higgs boson decaying into a $\mathrm{b\bar{b}}$ pair and the other decaying into  $\gamma\gamma$,  $\tau\tau$,  $\mathrm{b\bar{b}}$, or a pair of vector bosons, using the dataset collected at 13 TeV corresponding to 35.9 $\mathrm{fb^{-1}}$. While the predicted cross-section of HH production according to the SM is too low to be probed by the used dataset, several BSM models, such as the two-Higgs-doublet model \cite{bib:2HDM} and its realization
within the minimal supersymmetric standard model~\cite{bib:MSSM1, bib:MSSM2} predict an enhancement in the HH production that could be probed with the available dataset. A combination of the measurements in the different channels is done, leading to the setting of an upper limit on the production cross-section at 95\% confidence level (CL) observed (expected) to be 21.8 (12.4) times the SM expectation. A search for heavy narrow width resonances (X) decaying to Higgs boson pairs is also performed in the mass range 250 < $\mathrm{M_{X}}$ < 3000 GeV. No evidence of a signal is observed, and upper limits are set on the resonance production
cross-section. The limits set are significant improvements upon the previous limits set using the dataset collected during the LHC Run 1. Figure~\ref{fig:hh} shows the limits set on the HH production.

\subsection{SM Measurements}
\label{subsec:standard_model}

\begin{figure}[htbp]
\begin{center}
\includegraphics[width=12cm,clip]{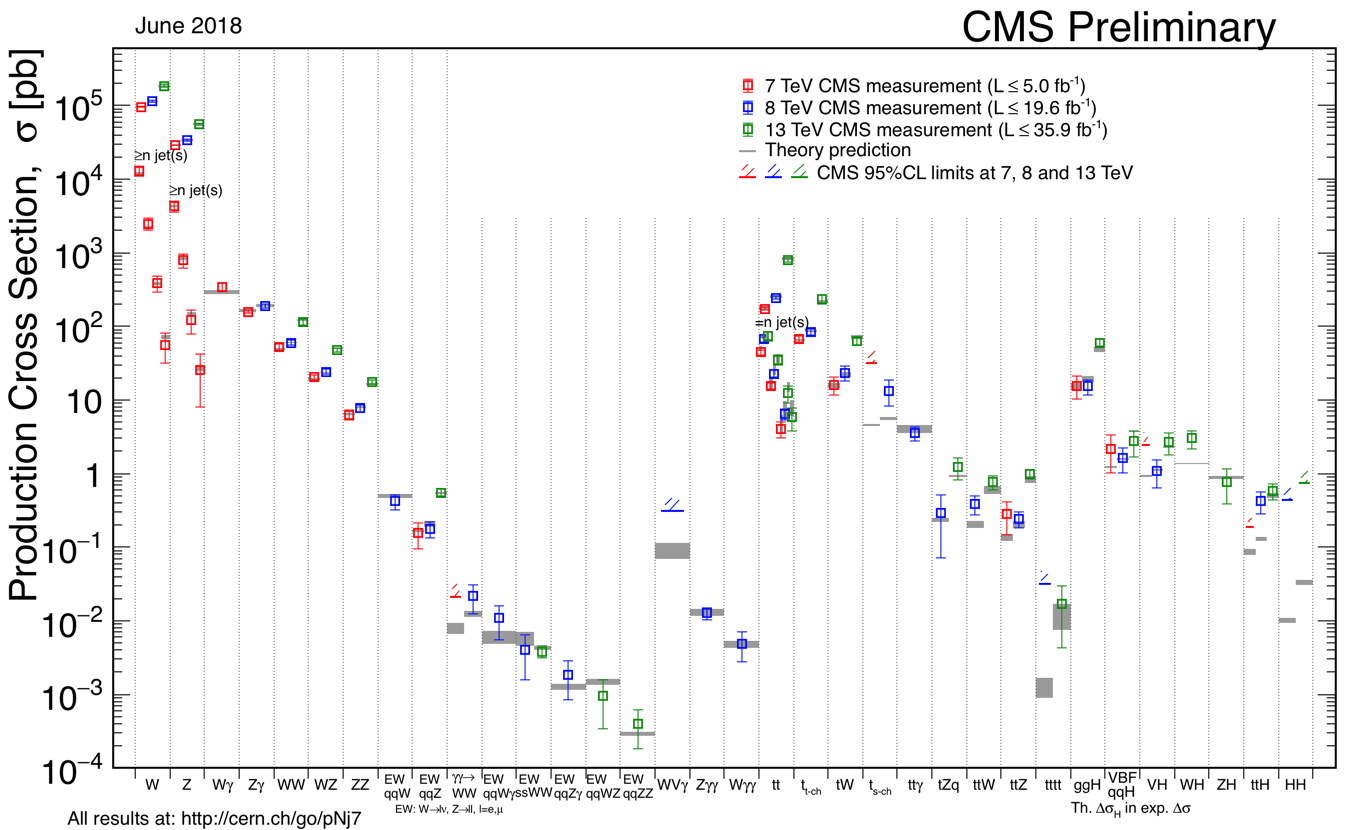}
\caption{Summary of the cross-section measurements of SM processes \cite{bib:cms_smp_summary}.}
\label{fig:sm_crosssections}       
\end{center}
\end{figure}

The SM has been very successful in describing the fundamental particles and their interactions. Figure~\ref{fig:sm_crosssections} shows a comparison of the CMS measured cross-section of several processes compared to the prediction of the cross-section according to SM. The measurements are found to be consistent with the SM predictions, displaying the remarkable success of SM.

Several analyses at CMS are dedicated to precisions measurements of the standard model, to probe the SM and to further improve our understanding of the particles and their interactions. Two such measurements,  of the WZ productions and of the production of the W boson in association with charm quarks, are described in the following sections.

\subsubsection{Measurement of WZ Production Cross-Section}

A measurement of the WZ diboson production cross-section has been performed in the leptonic decay channel \cite{bib:cms_wz}, targeting the final state with three leptons (muons or electrons) and a neutrino. This is the first such measurement with the full dataset collected during 2016 at 13 TeV center-of-mass energy. Measurements of the inclusive cross-section, charge asymmetry and the differential cross-section with respect to the mass of the WZ system, the transverse momentum of the Z boson and the transverse momentum of the leading jet have been made. The measured value of the inclusive cross-section is $\sigma_{tot}(pp\rightarrow WZ)$ = $\mathrm{ 48.09^{+1.00}_{-0.96} (stat) ^{+0.44}_{-0.37} (theo) ^{+2.39}_{-2.17} (syst) \pm 1.39 (lumi)}$ pb, for a total uncertainty of -2.78 and +2.98 pb. The precise measurements made are found to be compatible with the SM prediction at next-to-next-to-leading-order calculations.

\begin{figure}[htbp]
\begin{center}
\includegraphics[width=12.5cm,clip]{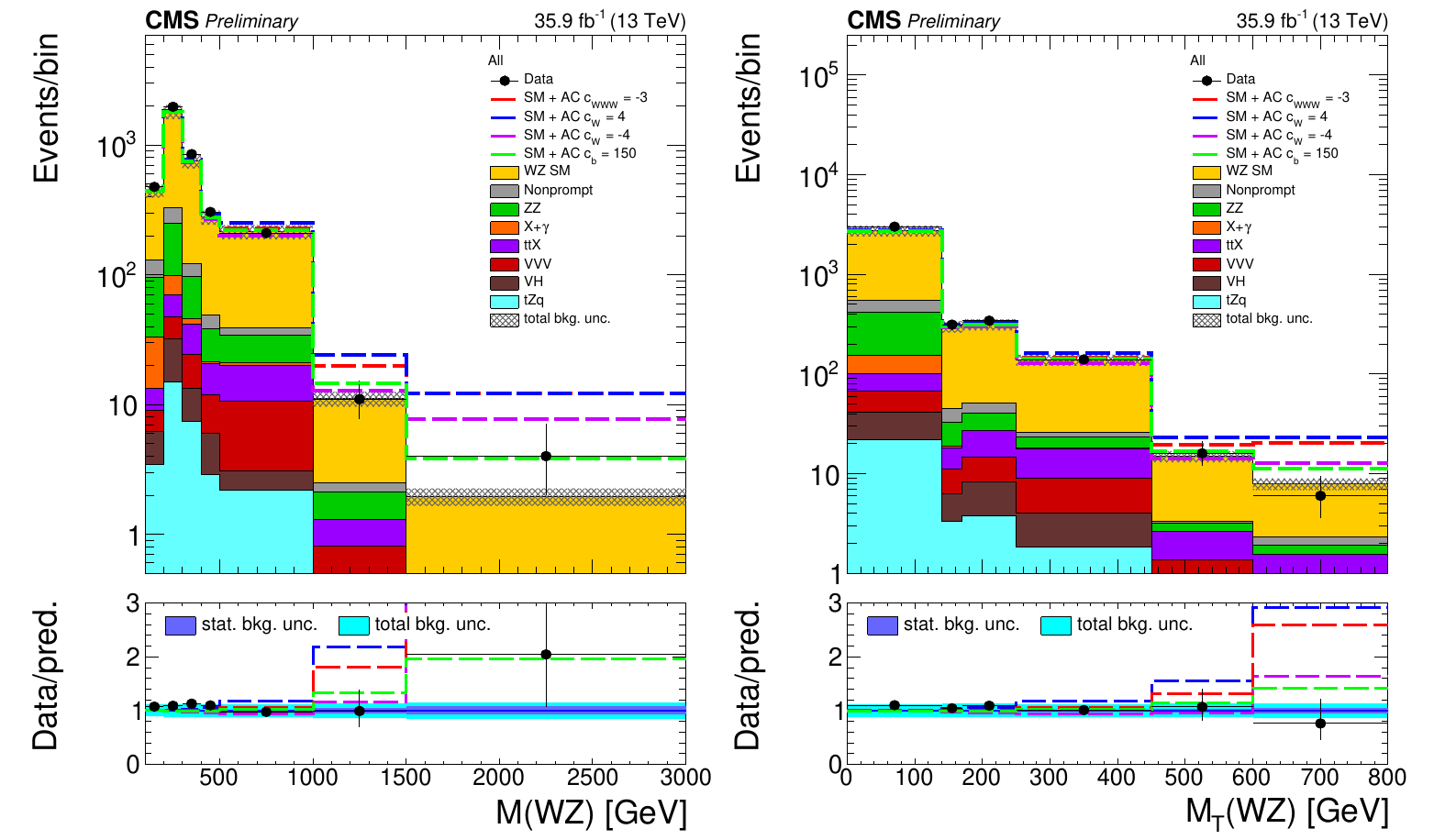}
\caption{Distributions of discriminant observables in the anomalous couplings searches. The invariant mass of the three lepton plus the missing transverse momentum system, representing the WZ invariant mass under the assumption that the momentum of the neutrino from the W boson decay is the same as the missing transverse momentum as described in Ref.~\cite{bib:cms_wz}, is shown on the left and the transverse mass of the same configuration is shown on the right. The dashed lines represent the total yields that would be expected from the sum of the SM processes with the total WZ yields modified according to their correspondence to the given values of the associated anomalous coupling parameters. The~SM prediction for the WZ process is obtained from the aTGC simulated sample with the anomalous couplings set to 0 \cite{bib:cms_wz}.}
\label{fig:wz_results}       
\end{center}
\end{figure}

The WZ process probed is sensitive to anomalous gauge couplings. An effective field theory (EFT)-based approach is used to set limits on possible anomalous gauge coupling. Constraints on anomalous triple gauge couplings are derived via a binned maximum likelihood fit to the WZ invariant mass variable, the distributions are shown in Figure~\ref{fig:wz_results}.

The parametrization of the additional terms for the anomalous triple gauge coupling (aTGC) is 

$\delta\mathcal{L}_{AC} = c_{www}Tr[W_{\mu\nu}W^{\nu\rho}W_{\rho}^{\mu}] + c_{w}(D_{\mu}H)^{\dagger}W^{\mu\nu}D_{\nu}H + c_{b}(D_{\mu}H)^{\dagger}B^{\mu\nu}D_{\nu}H$,
\\ where $W_{\mu\nu}$, $B_{\mu\nu}$ are the field strengths associated with the SM electroweak bosons and $H$ is the SM Higgs field \cite{bib:cms_wz}. Values predicted by the SM are $c_{w}$ = $c_{www}$ = $c_{b}$ = 0.  No significant deviation with respect to the SM is observed. The limits on the terms for anomalous gauge coupling are presented in Table~\ref{tab:wz_agc_limits}. These are the most stringent limit on the anomalous gauge coupling.

\begin{table}[htbp]
\centering
\caption{Expected and observed 1-D confidence intervals at 95\% confidence level for each of
the considered anomalous coupling parameters \cite{bib:cms_wz}.}
\label{tab:wz_agc_limits}       
\begin{tabular}{c|c|c}
\hline
\textbf{Parameter} & \textbf{95\% CI (Expected)} & \textbf{95\% CI (Observed)}  \\
\hline
$c_{w}/\Lambda^{2}$ & $-$13.3, 2.0] &  [$-$4.1, 1.1]  \\[0.5ex]
$c_{www}/\Lambda^{2}$ & [$-$1.8, 1.9]  & [$-$2.0, 2.1] \\[0.5ex]
$c_{b}/\Lambda^{2}$ & [$-$130, 170] &  [$-$100, 160] \\
\hline
\end{tabular}
\end{table}

\subsubsection{Measurement of W+c Differential Cross-Section}

\begin{figure}[htbp]
\begin{center}
\includegraphics[width=7cm,clip]{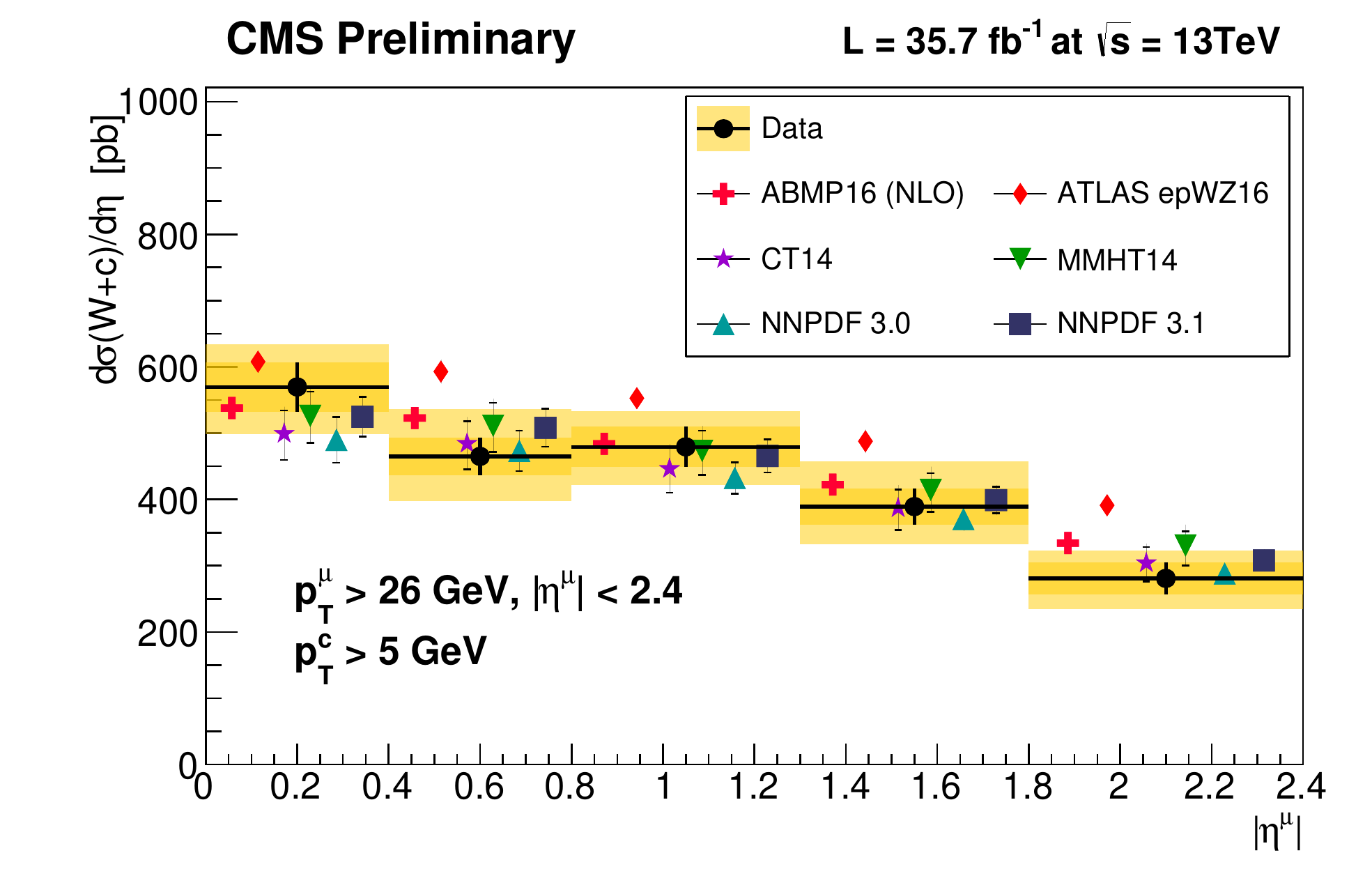}
\includegraphics[width=7cm,clip]{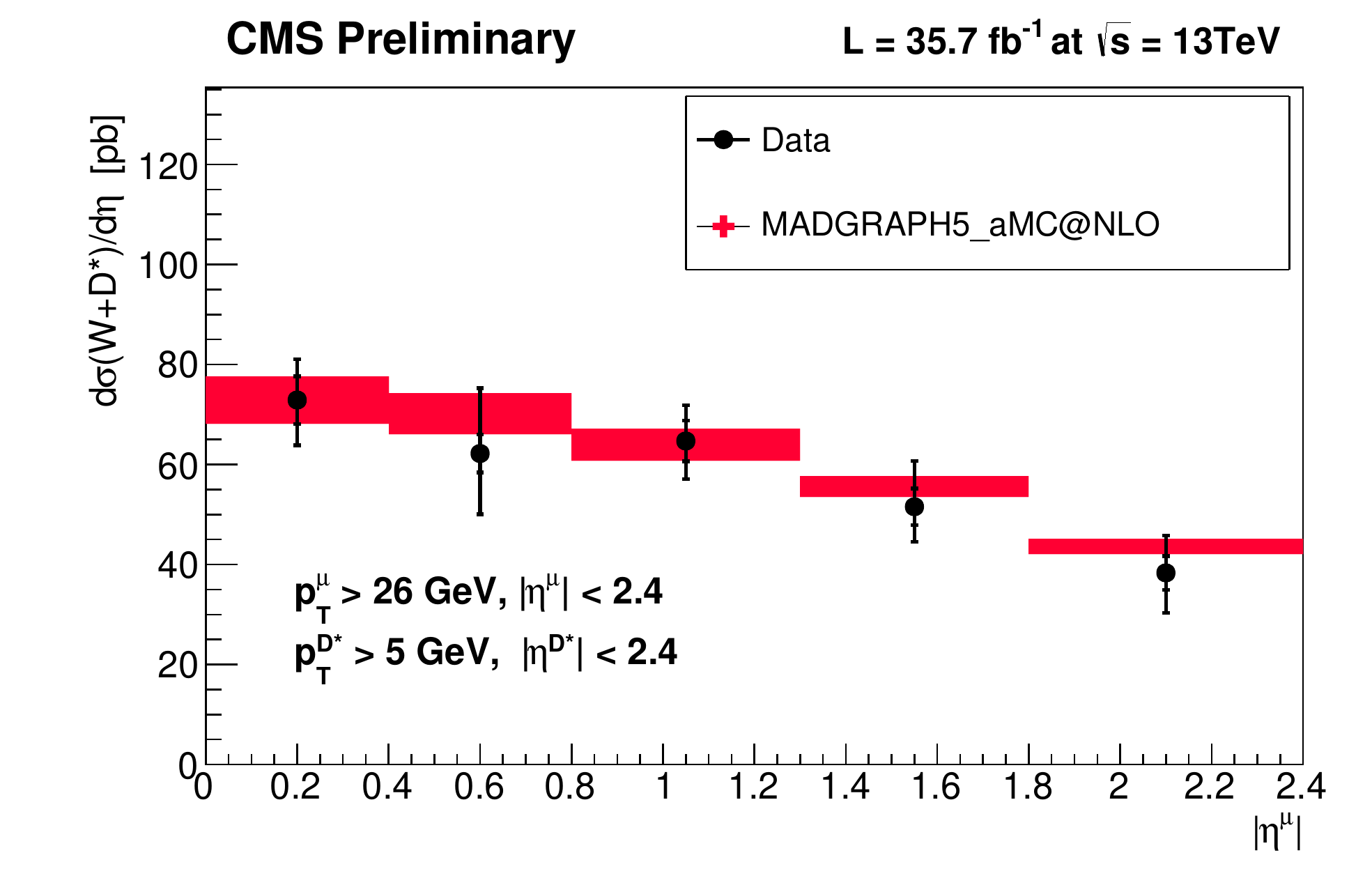}
\caption{Cross sections of $\sigma$(W+c) production at 13 TeV measured as a function of the pseudorapidity of the muon from the W boson decay (\textbf{left}), compared to the QCD predictions calculated with MCFM at NLO using different PDF sets. $\sigma$(W+D*) production cross sections presented as a function of the pseudorapidity of the muon from the W boson decay (\textbf{right}), compared to the predictions of the signal MC accounting for PDF uncertainties and scale variations, which are added in quadrature (shaded band) \cite{bib:cms_wc}.} 
\label{fig:wc_pdfs}       
\end{center}
\end{figure}

\begin{figure}[htbp]
\begin{center}
\includegraphics[width=6cm,clip]{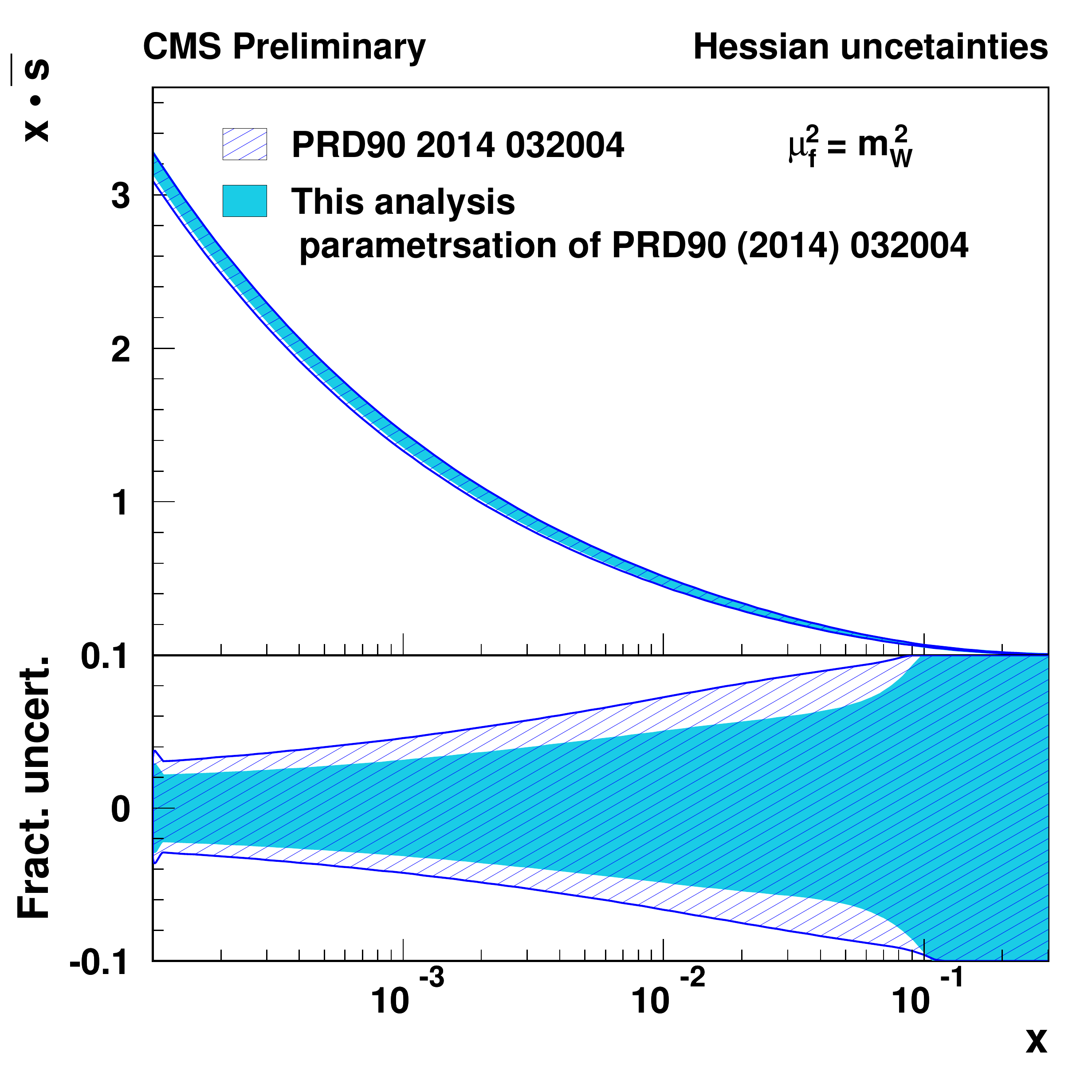}
\includegraphics[width=6cm,clip]{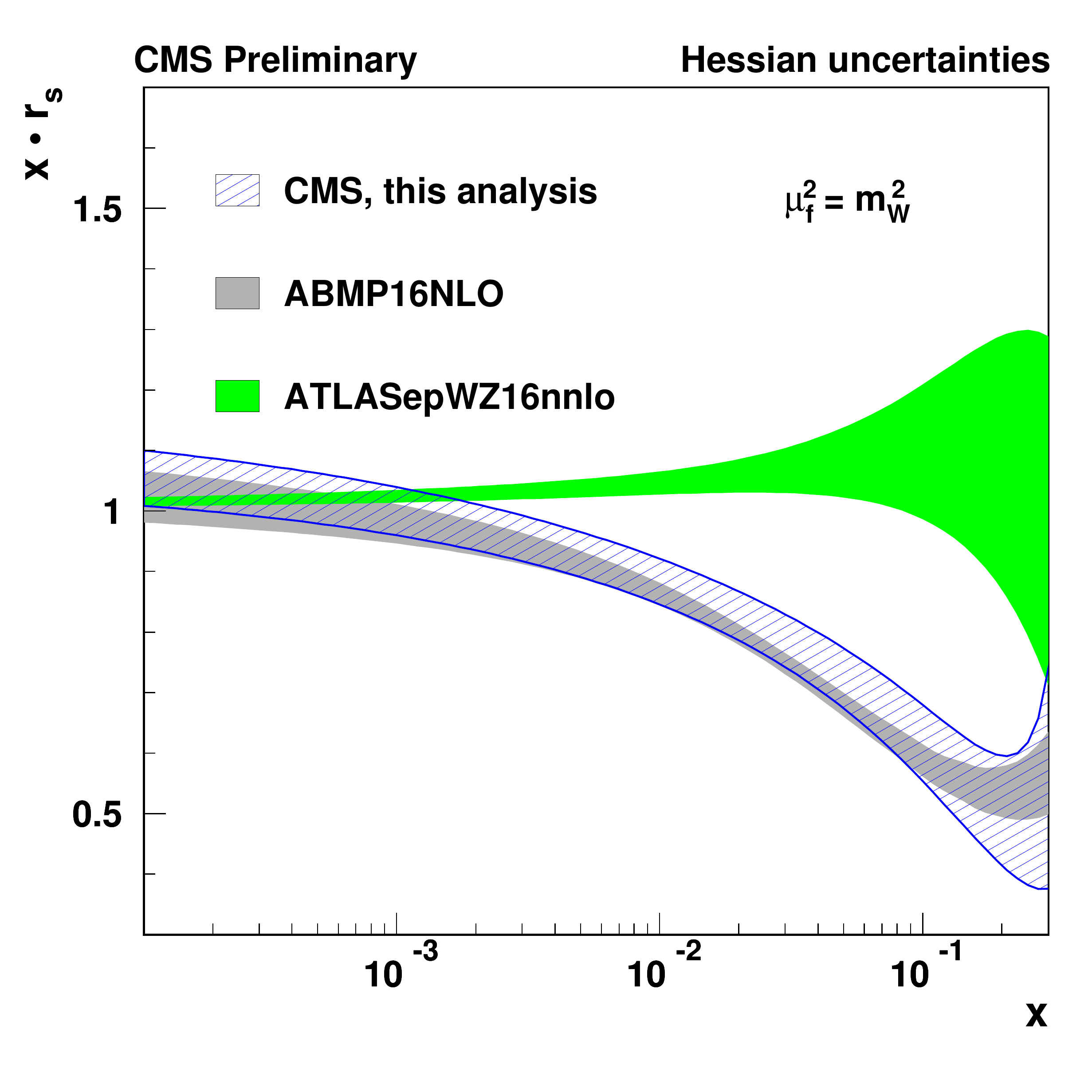}
\caption{ The distributions of s quark (\textbf{left upper} panel) in the proton and its relative uncertainty (\textbf{left lower} panel) as a function of $x$ at the factorization scale of $m_{W}^{2}$. The result of the current analysis (filled band) is compared to the result of a previous measurement \cite{bib:cms_wasym}.The strangeness suppression factor as a function of $x$  (\textbf{right}) at the factorization scale of $m_{W}^{2}$  compared to ABMP16nlo (dark shaded band) and ATLASepWZ16nnlo (light shaded band) PDFs \cite{bib:cms_wc}.}
\label{fig:wc_results}       
\end{center}
\end{figure}

The measurement of the differential cross-section of the associated production of a W boson with a charm quark (\textit{W+c}), 
in the $W\rightarrow\mu\nu$ decay channel of the W boson and the charm quark tagged with the $D^{*}(2010)^{\pm}$ production with $D^{*}(2010)^{\pm} \rightarrow D^{0} + \pi^{\pm}_{slow} \rightarrow K^{\mp} + \pi^{\pm} + \pi^{\pm}_{slow}$ , 
in terms of the pseudorapidity of the muon has been performed \cite{bib:cms_wc}. This measurement has been performed using the full proton-proton collision dataset at $\sqrt{s}$ = 13 TeV collected during 2016. The total cross-section of $\sigma(W+c) = 1026 \pm 31 (stat) ~^{+76}_{-72} (syst)$ pb is obtained. A precise measurement of the differential cross-section of the W+c productions is sensitive to the proton parton distribution functions (PDFs), particularly for the strange quark content. The measurements are compared to predictions according to different PDF sets, as shown in Figure~\ref{fig:wc_pdfs}. The potential of this measurement to improve the determination of the strange quark PDFs is shown in Figure~\ref{fig:wc_results}(left) using the procedure described in  Ref.~\cite{bib:cms_wc} with a comparison of the uncertainties obtained in the strange quark PDFs with respect to a previous measurement \cite{bib:cms_wasym}. It is seen that there is a reduction in the uncertainties on including this measurement. A measurement of the strangeness suppression factor is also performed (Figure~\ref{fig:wc_results}(right)) and it is found to be consistent with the predictions from the ABMP16NLO PDF set \cite{bib:abmp_pdf} and not with the ATLASepWZ16nnlo \cite{bib:wc_atlas}.

\subsection{Top Physics}
\label{subsec:top}

An important measurement of the single (anti) top quark production cross-section has been performed and is presented below.

\subsubsection{Measurement of Single (Anti) Top Quark Production Cross-Section and Ratio}

\begin{figure}[htbp]
\begin{center}
\includegraphics[width=7.8cm,clip]{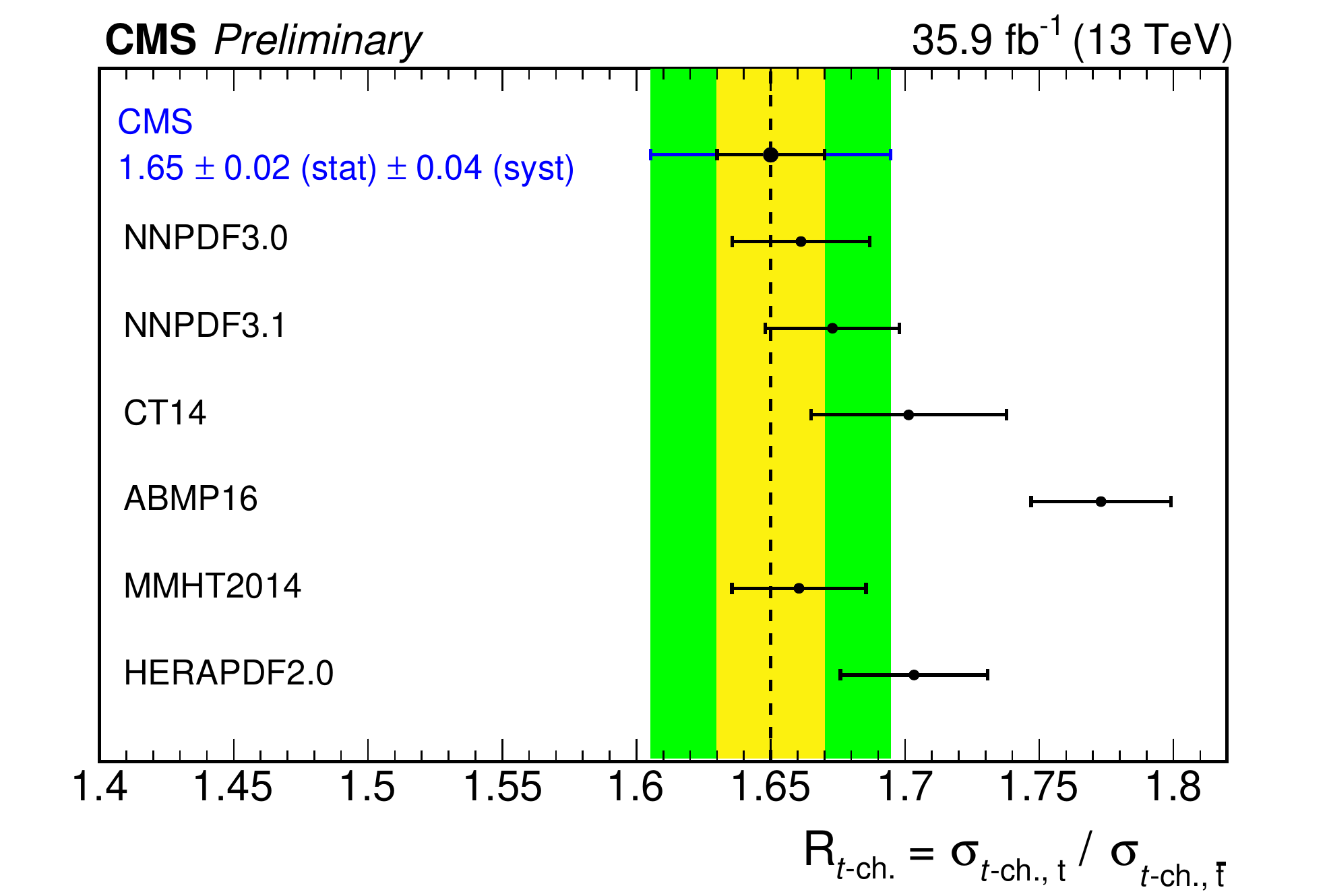}
\caption{Comparison of the measured $R_{t-ch}$ (dotted line) with the prediction from different PDF sets \cite{bib:cms_topcsn}.}
\label{fig:cms_top}       
\end{center}
\end{figure}

The single top and anti-top quark production cross sections have been measured \cite{bib:cms_topcsn}, corresponding to the t-channel production of the single (anti) top quark, in the leptonic channel with at least one electron or muon in the final state using the full proton-proton collision dataset at $\sqrt{s}$ = 13 TeV collected during 2016.  The total cross-section for the production of single top quarks or antiquarks is measured to be
$219.0 \pm 1.5 (stat) \pm 33.0 (syst)$ pb. A measurement of the ratio of the cross sections has been performed, and this very precise measurement can already probe the proton PDFs, as shown in the comparison of the CMS measurement with the predictions from the different PDF sets in Figure~\ref{fig:cms_top}. The~cross-section ratio is measured to be $R_{t-ch} = 1.65 \pm 0.02 (stat) \pm 0.04 (syst)$. The~measured ratio is used to obtain a direct measurement of CKM matrix element $V_{tb}$ to probe for any deviation from the expected value, and the measurement made is  $1.00 \pm 0.05 (exp) \pm 0.02 (theo)$ and is consistent with the SM expectation.  

\subsection{Flavor Physics}
\label{subsec:flavour}

Several important measurements pertaining to flavor physics have been performed at CMS. The~first observation of the resolved $\chi_{b}$(3P) states is presented below.

\subsubsection{Observation of $\chi_{b}$(3P) States}

\begin{figure}[htbp]
\begin{center}
\includegraphics[width=6cm,clip]{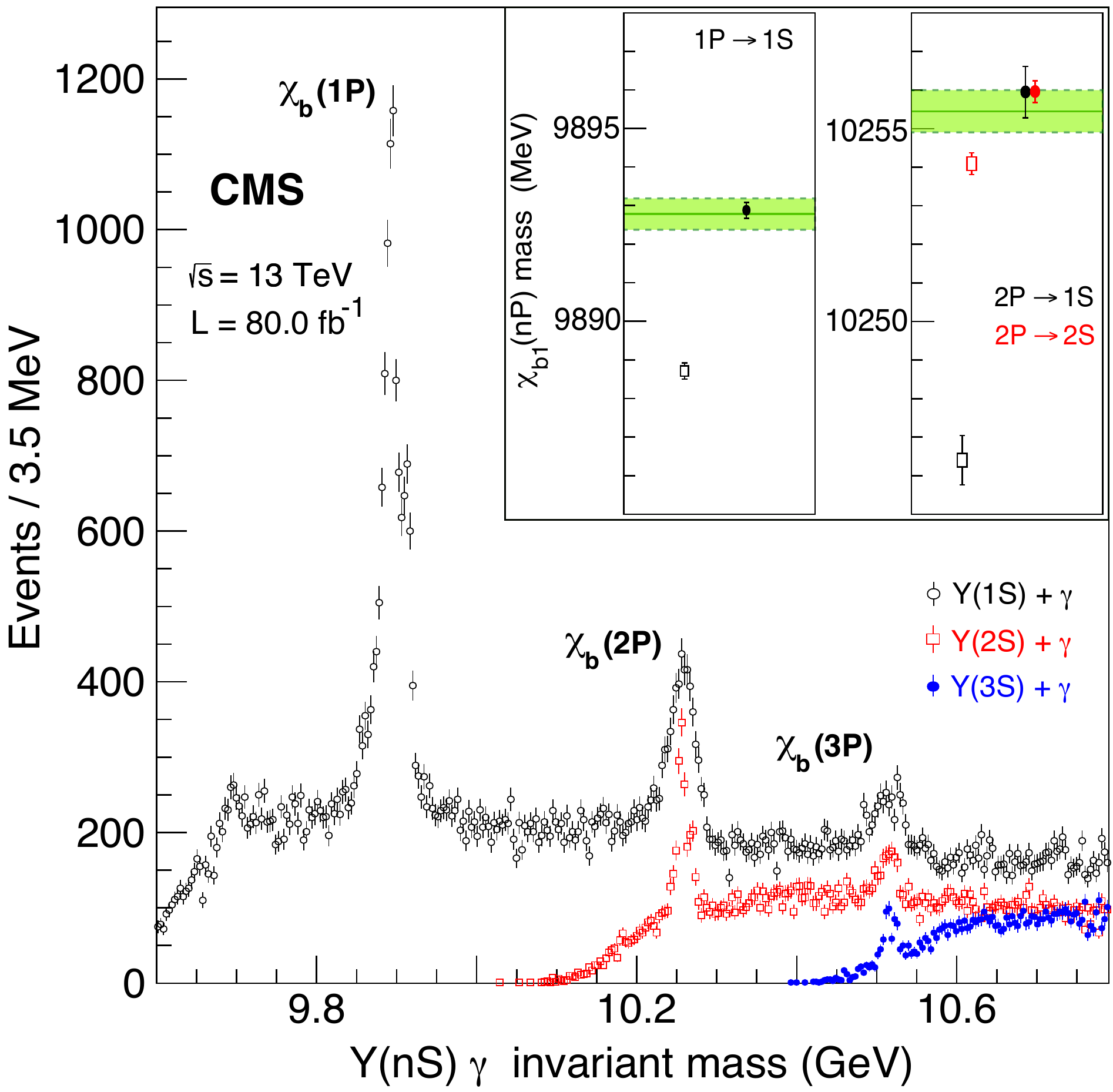}
\includegraphics[width=8cm,clip]{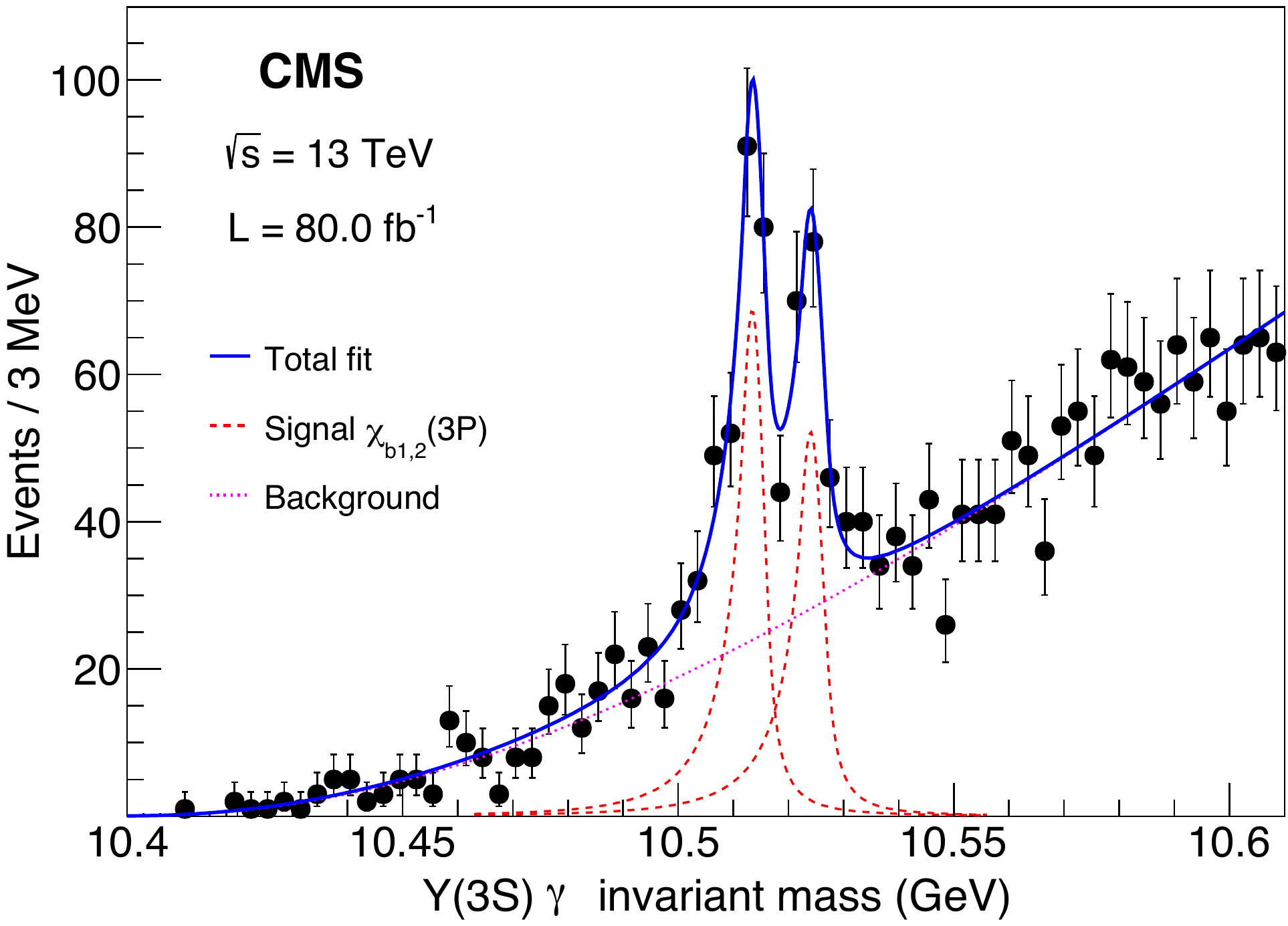}
\caption{ Invariant mass of the  $\Upsilon(nS)\gamma$ system on the left, with a peak corresponding to $\chi_{b}$(3P), and on the right is the distribution of the invariant mass of the  $\Upsilon(3S)\gamma$ system around the $\chi_{b}$(3P) mass \cite{bib:cms_chib}.}
\label{fig:cms_chib}       
\end{center}
\end{figure}

Measurements of $\chi_{b}$(3P) are performed through the radiative decay 
$\chi_{b}(3P)\rightarrow \Upsilon(3S)\gamma \rightarrow \mu\mu\gamma$ with the photon
reconstructed through conversions to $e^+e^-$ pairs \cite{bib:cms_chib}.
This measurement is performed using the full proton collision datasets at 13 TeV collected during 2015 through to 2017, corresponding to an integrated luminosity of  80 $\mathrm{fb^{-1}}$. Figure~\ref{fig:cms_chib} shows the invariant mass of the  $\Upsilon(nS)\gamma$ system on the left, with a peak corresponding to $\chi_{b}$(3P), and on the right is the distribution of the invariant mass of the  $\Upsilon(3S)\gamma$ system around the $\chi_{b}$(3P) mass.

This is the first experimental observation of the individually resolved states of $\chi_{b1}$(3P) and $\chi_{b2}$(3P) corresponding to \textit{J} = 1, 2. 
The mass of the two states have been found to be \mbox{$10,513.42 \pm 0.41 (stat) \pm 0.18 (syst)$} MeV and \mbox{$10,524.02 \pm 0.57 (stat) \pm 0.18 (syst)$} MeV  , with a mass difference of $\Delta M = 10.60 \pm 0.64 (stat) \pm 0.17 (syst)$ MeV between the two states. These measured values can be compared to the predictions of theoretical calculations  \cite{bib:theory_chib1, bib:theory_chib2, bib:theory_chib3, bib:theory_chib4, bib:theory_chib5, bib:theory_chib6, bib:theory_chib7, bib:theory_chib8, bib:theory_chib9, bib:theory_chib10, bib:theory_chib11, bib:theory_chib12}. This measurement favors the predictions for the $\Delta M$ range from non-perturbative QCD range from 8 to 18 MeV, while disfavoring the prediction of  $-$2 MeV for coupling with the open-beauty threshold.

\subsection{Searches for Physics Beyond the Standard Model (BSM)}
\label{subsec:bsm}

CMS has an extensive search program for BSM, including several dedicated searches for supersymmetry, Dark Matter searches and searches for exotic models such as excited leptons and leptoquarks among others. Previously, unexplored final states are also being probed at CMS, such as those with emerging jets. Selected recent results from CMS searches for BSM physics are presented~below.

An extensive list of searches conducted by CMS can be found in Ref.~\cite{bib:cms_public_web}.

\subsubsection{Search for Dark Matter}

\begin{figure}[htbp]
\begin{center}
\includegraphics[width=7cm,clip]{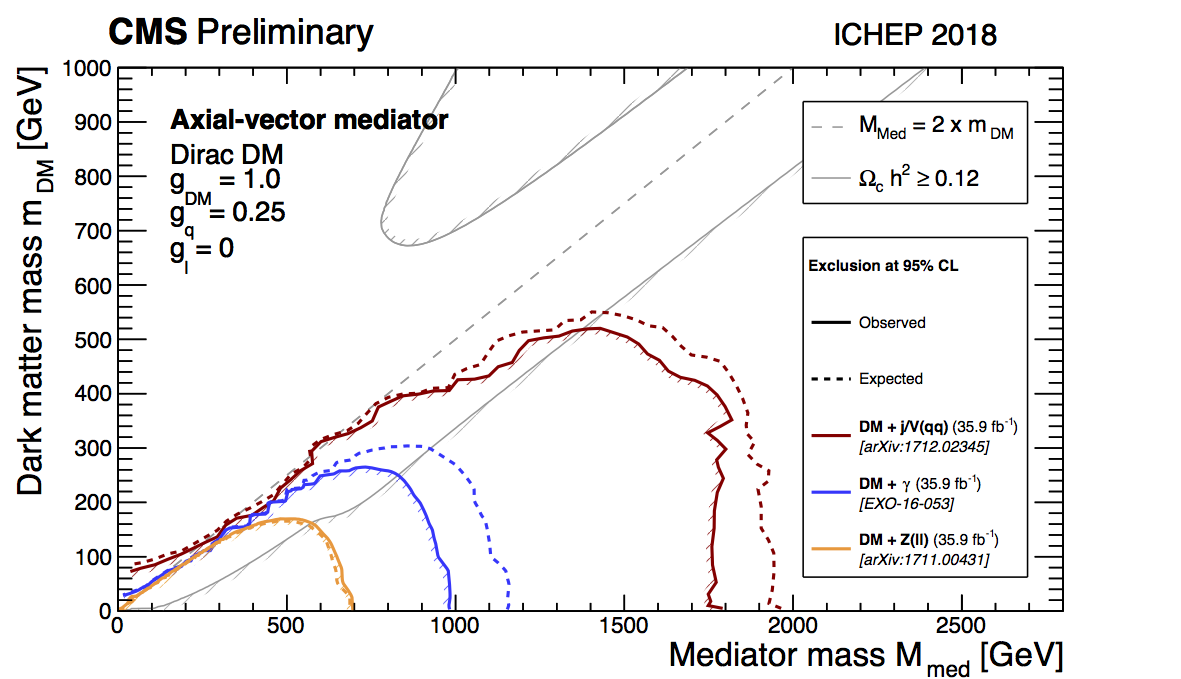}
\includegraphics[width=7cm,clip]{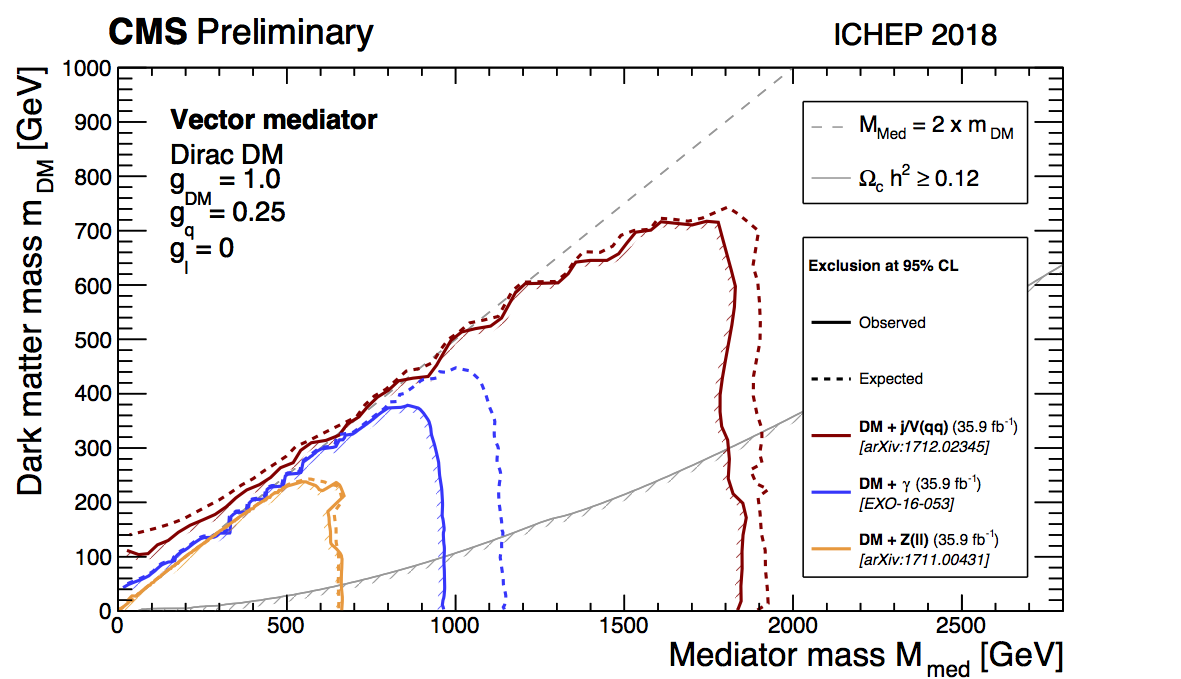}
\caption{ 95\% CL observed and expected exclusion regions in $m_{Med}-m_{DM}$ plane for different MET-based DM searches from CMS in the lepto-phobic Axial-vector model (\textbf{left}) and lepto-phobic Vector model (\textbf{right}) \cite{bib:cms_darkmatter_summary}.}
\label{fig:cms_dm}       
\end{center}
\end{figure}

Dark matter searches exemplify the rigorous approach at CMS that explores  a variety of different final states.
Dedicated analyses are performed in several different final states, particularly in final states with large missing transverse energy, corresponding to the dark matter associated particles, and another physics object, from initial state radiation. 
Several searches in different final states have been performed with 13 TeV datasets, including searches in the 
mono-jet \cite{bib:cms_darkmatter_monojet} final state, the~mono-photon final state \cite{bib:cms_darkmatter_monophoton} and the mono-Z final state \cite{bib:cms_darkmatter_monoV}. Since no significant excesses have been observed so far, a combination of limits from the different search analyses is made and is presented in Figure \ref{fig:cms_dm}.

\subsubsection{Search for Emerging Jets}

\begin{figure}[htbp]
\begin{center}
\includegraphics[width=7.5cm,clip]{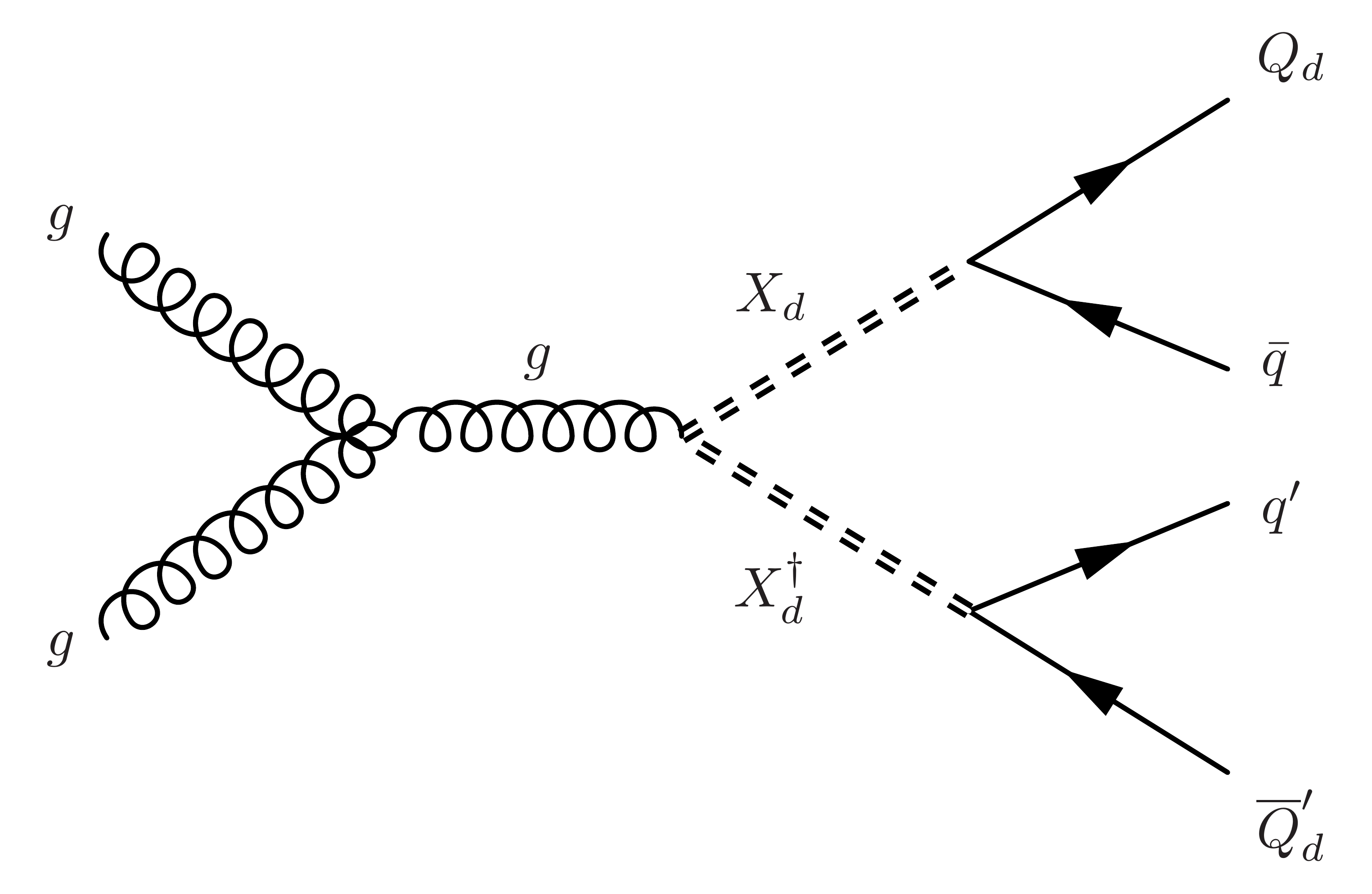}
\includegraphics[width=6cm,clip]{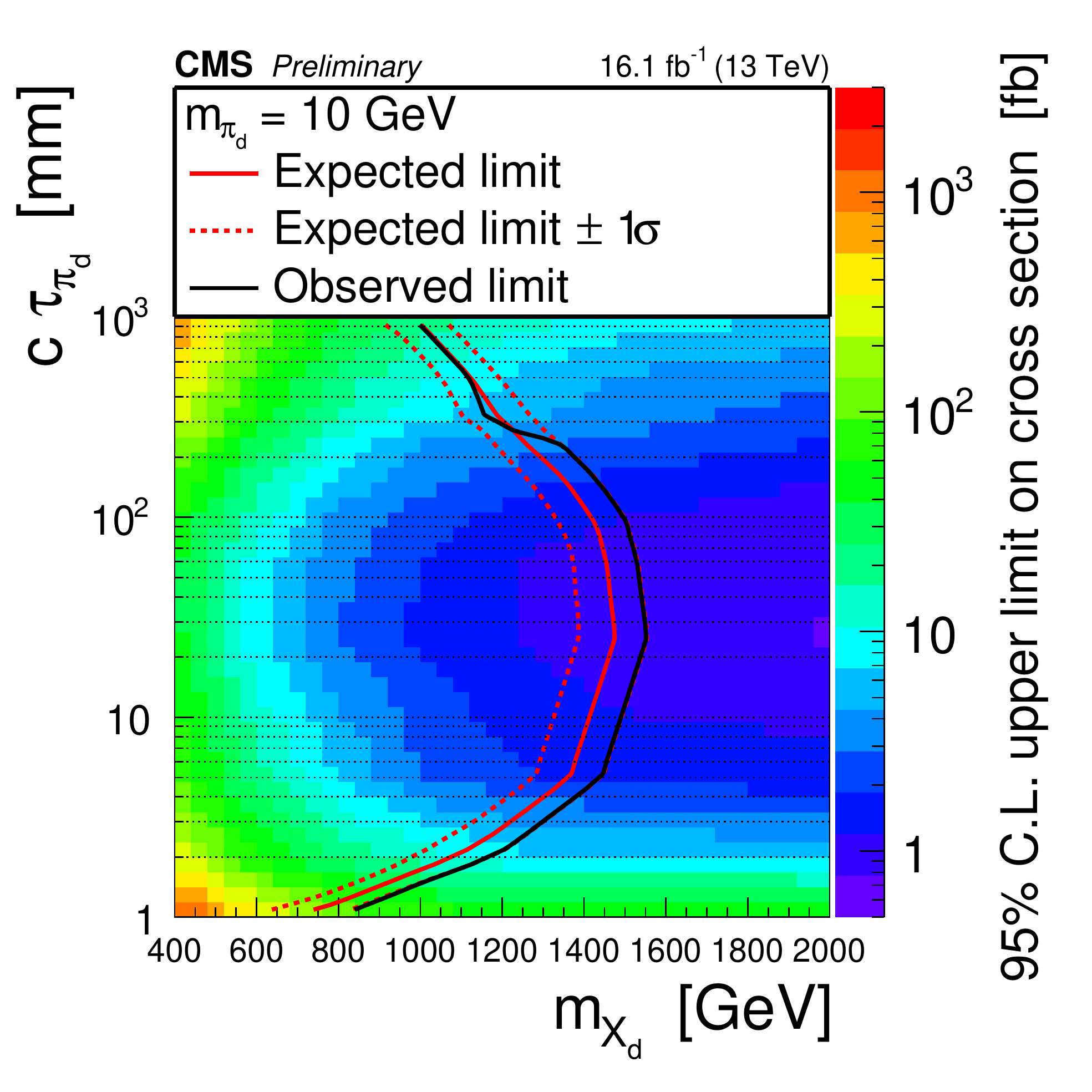}
\caption{ Feynman diagrams for pair production of mediator particles, with mediator decay to
a quark and a dark quark in the BSSW model \cite{bib:theory_emerging_jets_1, bib:theory_emerging_jets_2} via gluon fusion (\textbf{left}). Signal exclusion curves derived from theory-predicted cross sections and upper limits
at 95\% CL on the signal cross-section for models with dark pion mass of 10 GeV (\textbf{right}) \cite{bib:cms_emerging_jets}.}
\label{fig:emerging_jets}       
\end{center}
\end{figure}

The first dedicated search for emerging jets at the LHC has been performed by CMS \cite{bib:cms_emerging_jets}. This search is inspired by models containing dark baryons that could explain the dark matter observed in the universe \cite{bib:theory_emerging_jets_1, bib:theory_emerging_jets_2}. The search is for events consistent with the pair production of a heavy mediator particle that decays to a light quark and a new fermion, called a dark quark, using data corresponding to an integrated luminosity of 16.1 $\mathrm{fb^{-1}}$ from proton-proton collisions at $\sqrt{s}$ =13 TeV collected by the CMS experiment at the LHC in 2016. The dark quark gives rise to long-lived dark hadrons via a parton shower, with the resulting emerging jet containing
displaced vertices that are created by dark hadron decays to SM hadrons. Feynman diagrams for pair production of mediator particles, with mediator decay to
a quark and a dark quark in the BSSW model \cite{bib:theory_emerging_jets_1, bib:theory_emerging_jets_2} via gluon fusion is shown in Figure~\ref{fig:emerging_jets}(left).

A simple cut and count analysis strategy is used, with events selected with cut sets optimized with respect to model parameters. A major portion of phase space has been explored already, with no significant excess observed.
Mediator masses of 400 to 1250 GeV excluded for dark pion decay lengths of 5 to 225 mm. Signal exclusion curves derived from theory-predicted cross sections and upper limits
at 95\% CL on the signal cross-section are shown in Figure~\ref{fig:emerging_jets}(right) for models with a dark pion mass of 10 GeV.

\section{Conclusions}
\label{sec:conclusions}
The status of the CMS experiment along with a limited set of important recent results has been presented. A varied set of results covering measurements of the Higgs boson, precision SM measurements, analyses with top quarks and important flavor physics results has been presented.

The $\mathrm{t\bar{t}H}$ production mode of the Higgs boson was observed for the first time, and its signal strength is compatible with the standard model expectation within the uncertainties of the~measurement.

Several different searches for new physics BSM have also been presented, covering dark matter models, and new searches for models with emerging jets as signatures. The~dataset probed has been consistent with the SM prediction and no signs of new physics have been found yet. 

The CMS experiment has been performing excellently during the first two runs of the LHC and preparations for the upcoming third run of the LHC with proton-proton collisions at center-of-mass energy of 14 TeV and the following High-Luminosity LHC are progressing well. One can look forward to several more important analyses from CMS probing the Higgs boson and the SM, along with further searches for new physics using the large dataset collected that is yet to be analyzed completely, along with the larger datasets expected to be collected in the future.
 

\vspace{6pt}

\end{document}